\documentclass{revtex4}
\usepackage{graphicx}
\usepackage{fancyhdr}
\usepackage{amsmath}

\newcommand{\bra}[1]{\left< {#1} \,\right\vert}
\newcommand{\ket}[1]{\left\vert\, {#1} \, \right>}

\usepackage{color}

\begin{document}

\title{Physics at the ILC\\
 with focus mostly on Higgs physics}

%

\author{Keisuke Fujii}
\affiliation{High Energy Accelerator Research Organization (KEK), Tsukuba, Japan}

\begin{abstract}
Physics at the ILC is reviewed focusing mostly on Higgs physics.
It is emphasized that at the ILC it is possible to measure 
the $hZZ$ coupling totally model independently, which 
in turn allows model-independent normalization of various
branching ratio measurements and consequently the absolute
measurements of corresponding couplings.
Combining them with the measurements of the top Yukawa coupling
and the Higgs self-coupling at higher energies, the full ILC
program is shown to allow a precision test of the mass-coupling relation.
\end{abstract}

\maketitle

\thispagestyle{fancy}


\section{Introduction}

Let me begin my talk with the electroweak symmetry breaking and the mystery 
of something in the vacuum.
We all know that the success of the Standard Model (SM) of particle physics is a success
of gauge principle. 
We know that the transverse components of $W$ and $Z$ are gauge fields of the
electroweak (EW) gauge symmetry. 
Since the gauge symmetry forbids explicit mass terms for $W$ and $Z$, it must be
broken by {\it something condensed in the vacuum} which carries EW charges:
\begin{eqnarray}
\bra{0} I_3, Y \ket{0} \ne 0  \mbox{  while } \bra{0} I_3 + Y \ket{0}.
\end{eqnarray}
This "something" supplies three longitudinal modes of $W$ and $Z$:
\begin{eqnarray}
W_L^+, W_L^-, Z_L  \leftarrow \chi^+, \chi^-, \chi_3 \mbox{ : Goldstone modes.}
\end{eqnarray}
Since left- and right-handed matter fermions carry different EW charges, explicit mass
terms are also forbidden for matter fermions by the EW gauge symmetry.
Their masses have to be generated through their Yukawa interactions with some
weak-charged vacuum which compensates the EW-charge difference.
In the SM, the same "something" mixes the left- and right-handed matter fermions, 
consequently generating masses and inducing flavor-mixings among generations.
In order to form the Yukawa interaction terms, we need a complex doublet scalar field. 
The SM identifies three real components of the doublet with the Goldstone modes that supply
the longitudinal modes of $W$ and $Z$.
We need one more to form a complex doublet, which is the physical Higgs boson.
This SM symmetry breaking sector is the simplest and the most economical, but there is no reason for it.
The symmetry breaking sector (hear after cooled the Higgs sector) might be more complex.
We don't know whether the "something" is elementary or composite.
We know it's there in the vacuum with a vev of 246\,GeV. 
But other than that we didn't know almost anything about the "something" until July 4th, 2012.

Since the July 4th, the world has changed! 
The discovery of the 125\,GeV boson ($X(125)$) at the LHC could be called a quantum jump \cite{ref:LHChiggs}.
The $X(125) \to \gamma\gamma$ decay means $X$ is a neutral boson having a spin not
equal to 1 (Landau-Yang theorem).
We know that the 125\,GeV boson decays to $ZZ^*$ and $WW^*$, indicating the existence
of $XVV$ couplings, where $V=W/Z$, gauge bosons.
There is, however, no gauge coupling like $XVV$.
\begin{figure}[h]
\centering
\includegraphics[width=100mm]{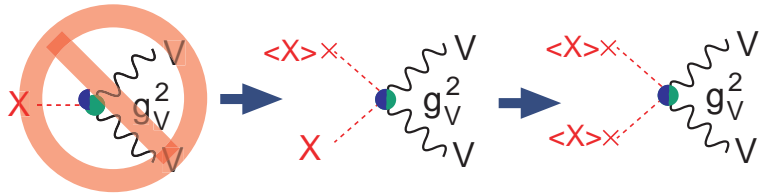}
\caption{The origin of $XVV$ coupling and its relation to the mass term of $V$.} \label{fig:xvv}
\end{figure}
There are only $XXVV$ and $XXV$, hence $XVV$ is most probably from
$XXVV$ with one $X$ replaced by its vacuum expectation value $\left< X \right> \ne 0$,
namely $\left<X\right>XVV$.
Then there must be $\left<X\right>\left<X\right>VV$, a mass term for $V$, meaning that
$X$ is at least part of the origin of the masses of $V=W/Z$.
This is a great step forward but we need to know whether $\left<X\right>$ saturates the
SM vev of 245\,GeV.
The observation of the $X \to ZZ^*$ decay means that $X$ can be produced via $e^+e^- \to Z^* \to ZX$.
\begin{figure}[h]
\centering
\includegraphics[width=85mm]{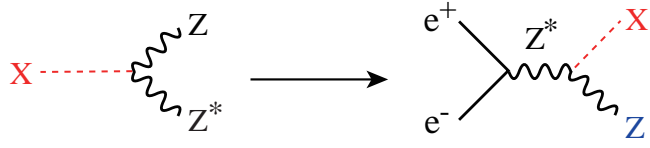}
\caption{$X \to ZZ^*$ decay and $e^+e^- \to ZX$ process.} \label{fig:xzzandzh}
\end{figure}
By the same token, $X \to WW^*$ means that $X$ can be produced via the $W$-fusion process:
$e^+e^- \to \nu\bar{\nu}X$.
So we now know that the major Higgs production processes in $e^+e^-$ collisions are indeed
available at the ILC, which can be regarded as a "no lose theorem" for the ILC.
The $125\,$GeV is the best place for the ILC, where variety of decay modes are accessible.
We need to check this $125\,$GeV boson in detail to see if it has indeed all the required
properties of the "something" in the vacuum.

\begin{figure}[h]
\centering
\includegraphics[width=85mm]{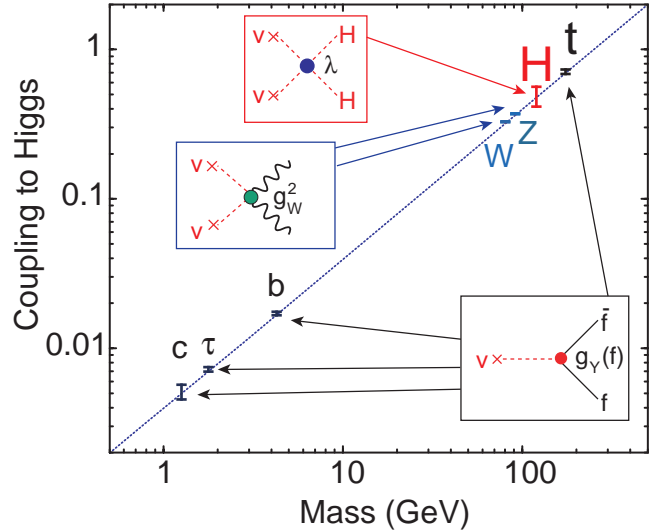}
\caption{Mass-coupling relation \cite{ref:acfa}.} \label{fig:mass-coupling1}
\end{figure}

The properties to measure are the mass, width, $J^{PC}$, gauge quantum numbers,
Yukawa couplings to various matter fermions, and its coupling to itself.
The key is to measure {\it the mass-coupling relation}.
If the 125\,GeV boson is the one to give masses to all the SM particles,
coupling should be proportional to mass as shown in Fig.\ref{fig:mass-coupling1}.
Any deviation from the straight line signals physics beyond the Standard Model (BSM).
The Higgs is a window to BSM physics.

Our mission is the bottom-up model-independent reconstruction of the electroweak symmetry 
breaking sector through the coupling measurements.
We need to determine the multiplet structure of the Higgs sector by answering questions like:
Is there an additional singlet or doublet or triplet?
What about the underlying dynamics? 
Is it weakly interacting or strongly interacting?
In other words, is the Higgs boson elementary or composite?
We should also try to investigate its possible relation to other questions of particle physics
such as dark matter, electroweak baryogenesis, neutrino masses, and inflation.
There are many possibilities to discuss and that's exactly why we are here in this meeting.
The July 4th was the opening of a new era which will last probably twenty years or more,
where a 500\,GeV linear collider such as the ILC will and must play the central role.

\section{Why 500\,GeV?}

There are three very well know thresholds.
\begin{figure}[h]
\centering
\includegraphics[width=140mm]{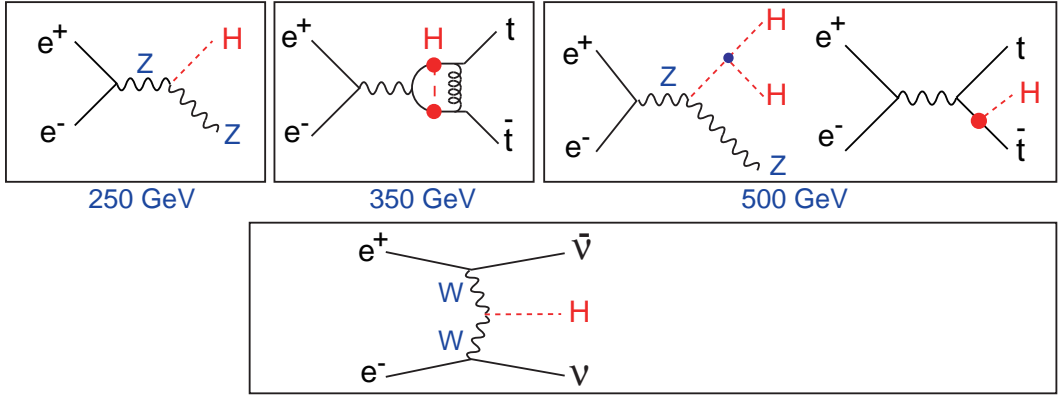}
\caption{Why 500\,GeV? The three thresholds.} \label{fig:the3thresholds}
\end{figure}
The first threshold is at around $\sqrt{s}=250\,$GeV, where the $e^+e^- \to Zh$ process will fully open.
We can use this process to measure the Higgs mass, width, and $J^{PC}$.
As we will see below, this process allows us to measure the $hZZ$ coupling in a completely
model-independent manner through the recoil mass measurement.
This is very important in extracting branching ratios for various decay modes such as
$h \to b\bar{b}, c\bar{c}, \tau\bar{\tau}, gg, WW^*, ZZ^*, \gamma\gamma$, as well as invisible decays.

The second threshold is at around $\sqrt{s}=350\,$GeV, which is the $t\bar{t}$ threshold.
Through the threshold scan, we can make a theoretically very clean measurement of the top quark mass, which can be translated into $m_t(\overline{MS})$ to an accuracy of $100\,$MeV.
The precision top mass measurement is, together with the precision Higgs mass measurement, very important from the view point of the stability of the electroweak vacuum \cite{ref:vacuumstability}.
The $t\bar{t}$ threshold also provides an opportunity to indirectly access the top Yukawa coupling through the Higgs exchange diagram as well as various $t\bar{t}$ bound state effects through the measurements of the forward-backward asymmetry and the top momentum, not to mention various form factor measurements to investigate possible anomaly in top-quark related couplings \cite{ref:dbd}.
It is also worth noting that the $\gamma\gamma$ collider option at this energy allows the double Higgs production: $\gamma\gamma \to hh$, which can be used to study the Higgs self-coupling \cite{ref:aahh}.
Notice also that at $\sqrt{s}=350\,$GeV and above, $e^+e^- \to \nu\bar{\nu}h$ process becomes sizable with which we can measure the $hWW$ coupling and accurately determine the total width, as we will see later.

The third threshold is at around $\sqrt{s}=500\,$GeV, where the production cross section for
$e^+e^- \to Zhh$ process attains its maximum, which allows us to access the Higgs self-coupling.
At $\sqrt{s}=500\,$GeV, another important process, $e^+e^- \to t\bar{t}h$, will also open
though the product cross section is much smaller than its maximum that happens at around
$\sqrt{s}=800\,$GeV. 
Nevertheless, as we will see, QCD threshold correction enhances the cross section and allows
us to measure the top Yukawa coupling with a reasonable precision concurrently with the self-coupling.

By covering $\sqrt{s}=250$ to $500\,$GeV, we can hence complete the mass-coupling plot.
This is why the first phase of the ILC project is designed to cover the energy up to $\sqrt{s}=500\,$GeV.

\section{ILC at 250\,GeV}

\begin{figure}[h]
\centering
\includegraphics[width=90mm]{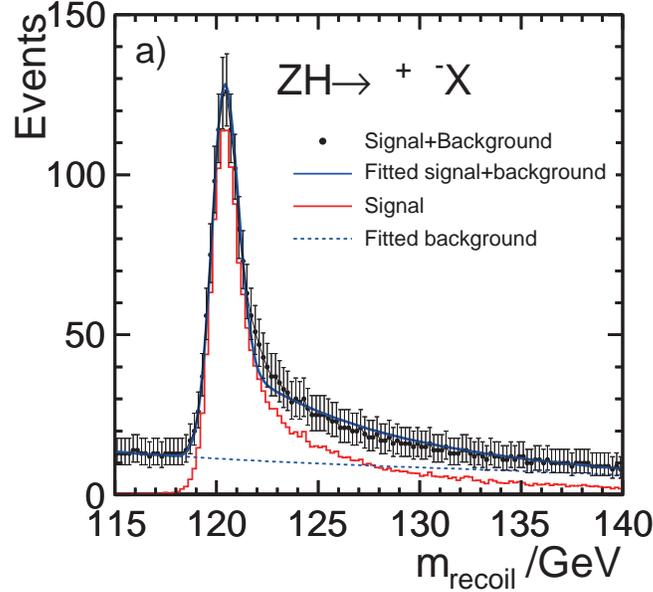}
\caption{Recoil mass distribution for the process: $e^+e^- \to Zh$ followed by $Z \to \mu^+\mu^-$ decay
for $m_h=120\,$GeV with $250\,$fb$^{-1}$ at $\sqrt{s}=250\,$GeV \cite{ref:ILDLoI}.} 
\label{fig:mrec}
\end{figure}

Let us now start with the first threshold at around $\sqrt{s}=250\,$GeV.
Perhaps the most important measurement at this energy is the recoil mass measurement
for the process: $e^+e^- \to Zh$ followed by $Z \to \ell^+\ell^- ~(\ell=e,\mu)$ decay.
Since the initial state 4-momentum is precisely known, we can calculate the
invariant mass of the system recoiling against the lepton pair from the $Z$ decay
by just measuring the momenta of the lepton pair:
\begin{eqnarray}
M_X^2 = \left(p_{CM} - (p_{\ell^+} + p_{\ell^-})\right)^2 .
\end{eqnarray}
Figure \ref{fig:mrec} shows the recoil mass distribution for a $m_h=120\,$GeV Higgs boson, with 
250\,fb$^{-1}$ at $\sqrt{s}=250\,$GeV.
You can see a very clean Higgs peak with small background.
Since we don't need to look at the Higgs decay at all, its invisible decay is also detectable.
This way, we can determine the Higgs mass to $\Delta m_h=30\,$MeV and the production cross section to $\Delta \sigma_{Zh} /\sigma_{Zh} = 2.5\,$\%, and limit the invisible branching ratio to $1\%$ at the $95\%$ confidence level.
This is the flagship measurement of the ILC at 250\,GeV that allows a model-independent absolute measurement of the $hZZ$ coupling\cite{ref:2012taa,ref:sid.loi}.\\

We can also use the $e^+e^- \to Zh$ process to measure various branching ratios for various Higgs decay modes.
This time we include $Z \to q\bar{q}$ and $\nu\bar{\nu}$ decays in our analysis to enhance the statistical precision. 
Notice, however, that what we can actually measure is {\it NOT} branching ratio ($BR$) itself but the cross section times branching ratio ($\sigma \times BR$).
Table \ref{tab:sigBR250} summarizes the expected precisions for the $\sigma\times BR$ measurements \cite{ref:BRs250, ref:BRtau}
\begin{table}[h]
\begin{center}
\caption{Expected relative errors for the $\sigma\times BR$ measurements at $\sqrt{s}=250\,$GeV with $250\,$fb$^{-1}$ for $m_h=120\,$GeV.}
\begin{tabular}{|c|c|c|c|}
\hline
process &decay mode & $\Delta \sigma BR/\sigma BR$ &  $\Delta BR/BR$ \\
\hline
Zh & $h \to b\bar{b}$ & 0.94\% & 2.7\% \\
\cline{2-4}
     & $h \to c\bar{c}$ & 6.5\% & 7.0\% \\
\cline{2-4}
     & $h \to gg$ & 8.0\%  & 8.4\% \\
\cline{2-4}
     & $h \to WW^*$ & 7.6\% & 8.0\% \\
\cline{2-4}
     & $h \to \tau\bar{\tau}$ & 3.4\% & 4.2\% \\
\cline{2-4}
     & $h \to ZZ^*$ & 25\% & 25\% \\
\cline{2-4}
     & $h \to \gamma\gamma$ & 23-30\%  & 23-30\% \\
\hline
\end{tabular}
\label{tab:sigBR250}
\end{center}
\end{table}
In order to extract $BR$ from $\sigma \times BR$, we need $\sigma$ from the recoil mass measurement,
hence the cross section error, $\Delta \sigma_{Zh}/\sigma_{Zh}=2.5\%$, eventually limits the BR measurements.
If we want to improve this, we need more data at $\sqrt{s}=250\,$GeV.
Notice here that "times two" luminosity upgrade is quite possible by increasing the number of bunches per train back to the original value of the reference design report \cite{ref:RDR}.\\

In order to extract couplings from branching ratios, we need the total width, since
the $hAA$ coupling squared is proportional to the partial width which is given by the
total width times the branching ratio:
\begin{eqnarray}
g_{hAA}^2 \propto \Gamma(h \to AA) = \Gamma_h \cdot BR(h \to AA).
\end{eqnarray}
Solving this for the total width, we can see that we need at least one partial width
and corresponding branching ratio to determine the total width:
\begin{eqnarray}
\Gamma_h = \Gamma(h \to AA) / BR(h \to AA) .
\end{eqnarray}
In principle, we can use $A=Z$ or $A=W$, for which we can measure both the $BR$s and the couplings.
In the first case, $A=Z$, we can determine $\Gamma(h \to ZZ^*)$ from the recoil mass measurement
and $BR(h \to ZZ^*)$ from the $\sigma_{Zh} \times BR(h \to ZZ^*)$ measurement together with the $\sigma_{Zh}$ measurement from the recoil mass.
This method, however, suffers from the low statistics due to the small branching ratio, $BR(h \to ZZ^*)=O(1\%)$,
A better way is to use $A=W$, where $BR(h \to WW^*)$ is subdominant and $\Gamma(h \to WW^*)$ can be determined by the $W$-fusion process: $e^+e^- \to \nu\bar{\nu}h$.
The measurement of the $W$-fusion process is, however, not easy at $\sqrt{s}=250\,$GeV since the cross section is small. Nevertheless, we can determine the total width to $\Delta \Gamma_h /\Gamma_h = 11\%$ with $250\,$fb$^{-1}$ \cite{ref:Durig}.
Since the $W$-fusion process becomes fully active at $\sqrt{s}=500\,$GeV, a much better measurement of the total width is possible there.
Let us then move on to the ILC at $\sqrt{s}=500\,$GeV.

\section{ILC at 500\,GeV}

At $\sqrt{s}=500\,$GeV, the $W$-fusion process $e^+e^- \to \nu\bar{\nu}h$ takes over the higgsstrahlung process: $e^+e^- \to Zh$.
We can use this $W$-fusion process for the $\sigma \times BR$ measurements as well as to determine the total width to $\Delta \Gamma_h / \Gamma_h = 6\%$.
Table \ref{tab:sigBR500} summarizes the $\sigma \times BR$ measurements for various modes.
\begin{table}[h]
\begin{center}
\caption{Expected relative errors for the $\sigma\times BR$ measurements at $\sqrt{s}=250\,$GeV with $250\,$fb$^{-1}$ and at $\sqrt{s}=500\,$GeV with $500\,$fb$^{-1}$ for $m_h=120\,$GeV and $(e^{-}, e^{+})=(-0.8, +0.3)$ beam polarization. The last column of the table shows the relative errors on branching ratios. Then numbers in the parentheses are as of $250\,$fb$^{-1}$ at $\sqrt{s}=250\,$GeV alone.}
\begin{tabular}{|l|r|r|r|r|}
   \hline
             & \multicolumn{3}{c|}{$\Delta (\sigma \cdot BR) / (\sigma \cdot BR)$} & $\Delta BR/BR$ \\
  \hline
   mode & $Zh$\,@\,250\,GeV & $Zh$\,@\,500\,GeV 
             & $\nu\bar{\nu}h$\,@\,500\,GeV 
             & combined \\
   \hline\hline
   $h \to b\bar{b}$       & 0.94\% & 1.6\%  & 0.60\%   & 2.6 (2.7)\% \\
   \hline
   $h \to c\bar{c}$        & 6.5\%  & 11\%    & 5.2\%     & 4.6 (7.0)\%  \\
   \hline
   $h \to gg$                & 8.0\%   & 13\%   & 5.0\%     & 4.8 (8.4)\%  \\
   \hline
   $h \to WW^*$          & 7.6\%   & 12.5\%  & 3.0\%   & 3.8 (8.0)\%  \\
   \hline
   $h \to \tau^+\tau^-$ & 3.4\%   & 4.6\%    & 11\%    &  3.6 (4.2)\%  \\
   \hline
\end{tabular}
\label{tab:sigBR500}
\end{center}
\end{table}
We can see that the $\sigma_{\nu\bar{\nu}h} \times BR(h \to b\bar{b})$ can be very accurately measured
to better than $1\%$ and the $\sigma_{\nu\bar{\nu}h} \times BR(h \to WW^*)$ to a reasonable precision with $500\,$fb$^{-1}$ at $\sqrt{s}=500\,$GeV.
The last column of the table shows the results of $\Delta BR / BR$ from the global analysis combining all the measurements including the total cross section measurement using the recoil mass at $\sqrt{s}=250\,$GeV. 
The numbers in the parentheses are with the $250\,$GeV data alone.
We can see that the $\Delta BR(h \to b\bar{b})/BR(h \to b\bar{b})$ is already limited by the recoil mass measurements.\\

\begin{figure}[h]
\centering
\includegraphics[height=55mm]{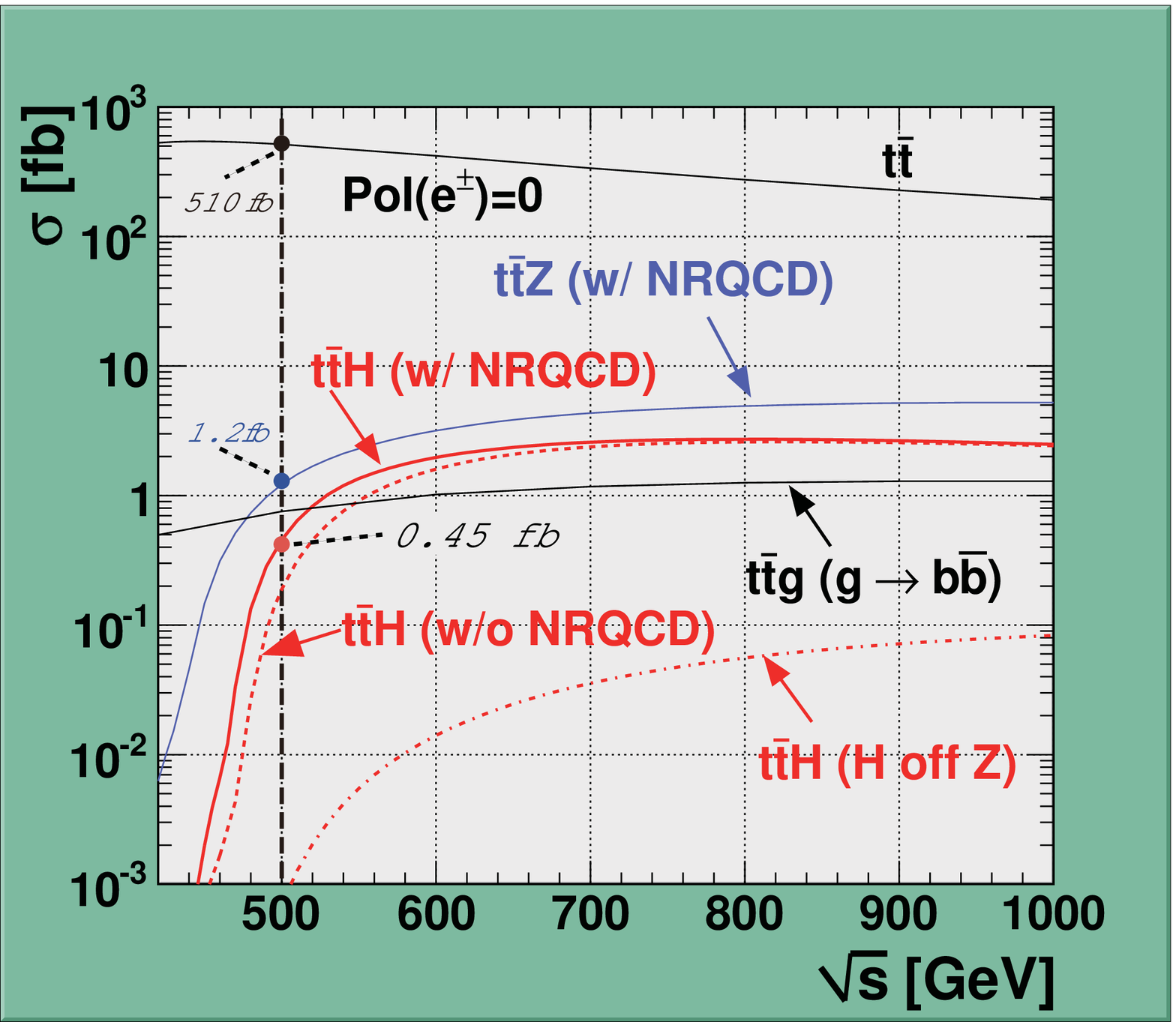}
\hspace{1cm}
\includegraphics[height=55mm]{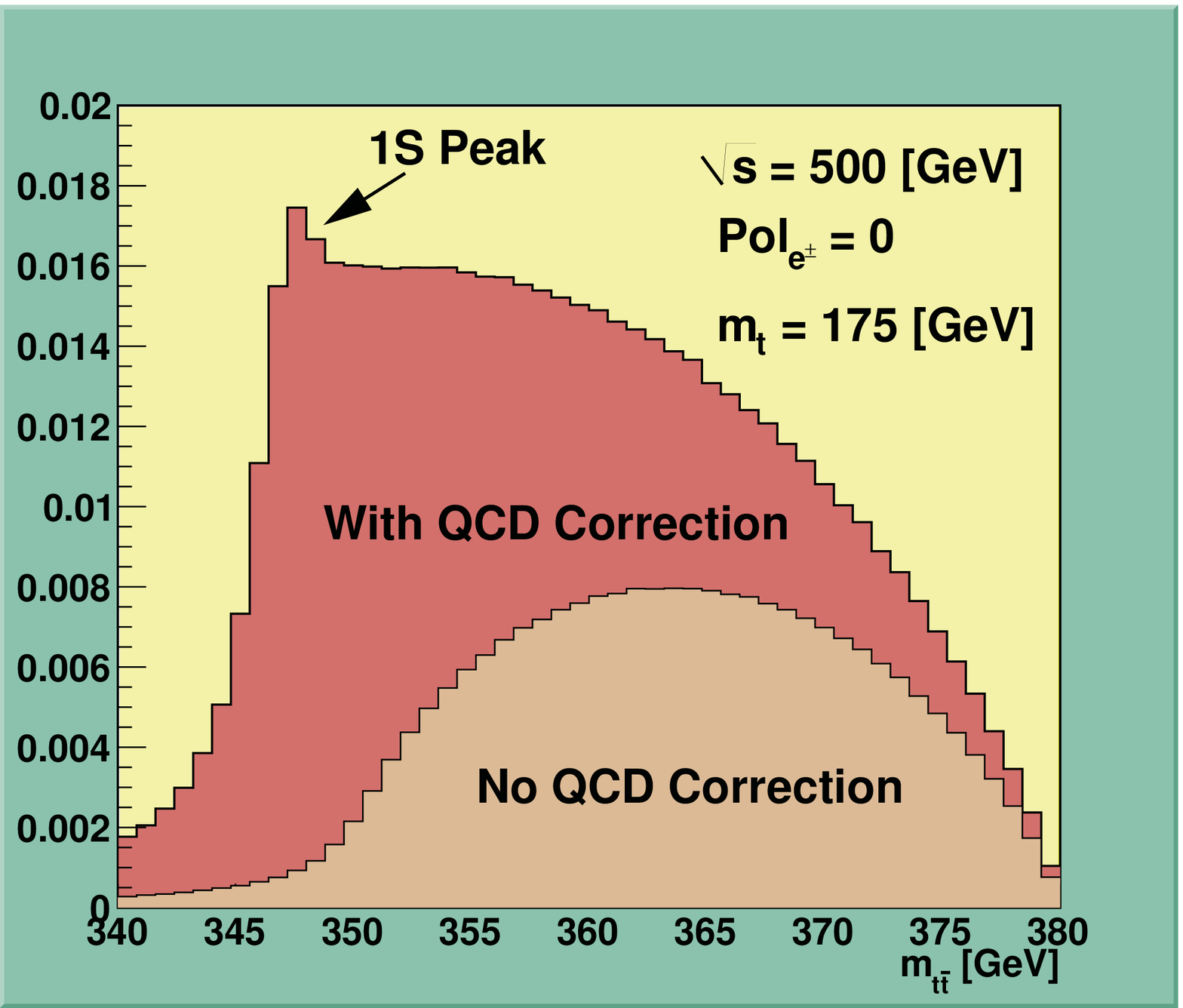}
\caption{Cross sections for the signal $t\bar{t}h$ process with and without the non-relativistic QCD (NRQCD) correction together with those for the background processes: $t\bar{t}Z, t\bar{t}g (g \to b\bar{b})$ and $t\bar{t}$ (left). The invariant mass distribution for the $t\bar{t}$ subsystem with and without the NRQCD correction (right).} 
\label{fig:sigtth}
\end{figure}

Perhaps more interesting than the branching ratio measurements is the measurement of the top Yukawa coupling using the $e^+e^- \to t\bar{t}h$ process, since it is the largest among matter fermions and not yet observed. 
Although the cross section maximum is reached at around $\sqrt{s}=800\,$GeV as seen in Fig.\ref{fig:sigtth}, 
the process is accessible already at $\sqrt{s}=500\,$GeV, thanks to the QCD bound-state effects (non-relativistic QCD correction) that enhance the cross section by a factor of two.
Since the background $h$-off-$Z$ diagram makes negligible contribution to the signal process, we can measure the top Yukawa coupling by simply counting the number of signal events.
The expected statistical precision for the top Yukawa coupling is then $\Delta g_Y(t) / g_Y(t) = 10\%$ with $1$ab$^{-1}$ at $\sqrt{s}=500\,$GeV \cite{ref:tth}.
Notice that if we go up by $20\,$GeV in the center of mass energy, the cross section doubles.
Moving up a little bit hence helps significantly.\\

\begin{figure}[h]
\centering
\includegraphics[height=60mm]{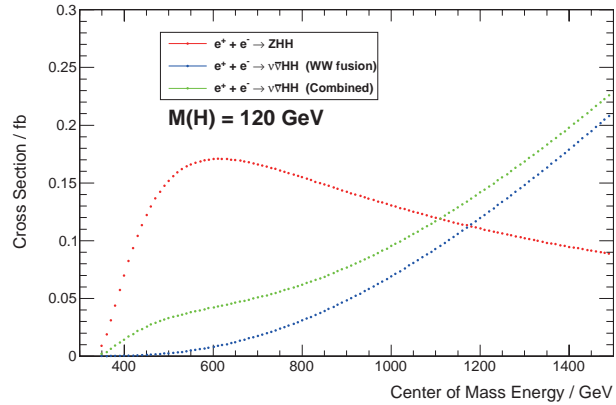}
\caption{Cross sections for the double Higgs production processes, $e^+e^- \to Zhh$ and $e^+e^- \to \nu\bar{\nu}hh$, as a function of $\sqrt{s}$ for $m_h=120\,$GeV.} 
\label{fig:sigzhh_vvhh}
\end{figure}

\begin{figure}[h]
\centering
\includegraphics[width=130mm]{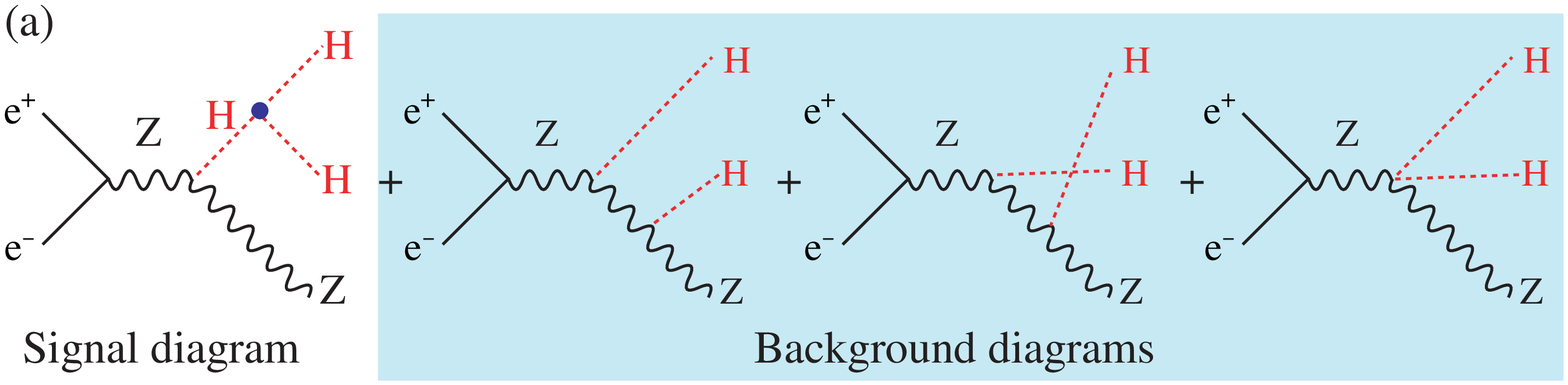}
\\~\\
\includegraphics[width=130mm]{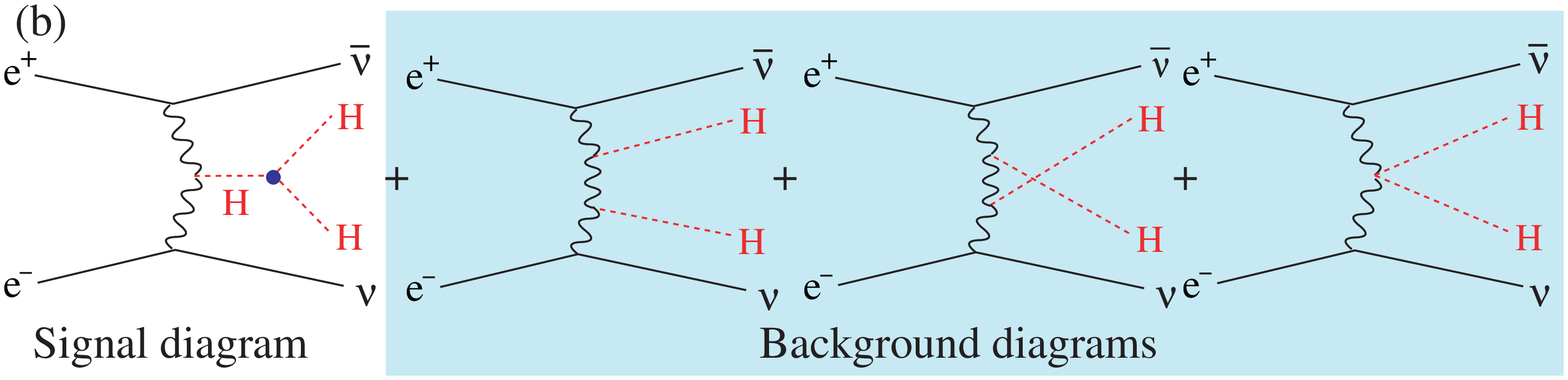}
\caption{Diagrams contributing to (a) $e^+e^- \to Zhh$ and (b) $e^+e^- \to \nu\bar{\nu}hh$.} 
\label{fig:hhdiagrams}
\end{figure}

\begin{figure}[h]
\centering
\includegraphics[height=50mm]{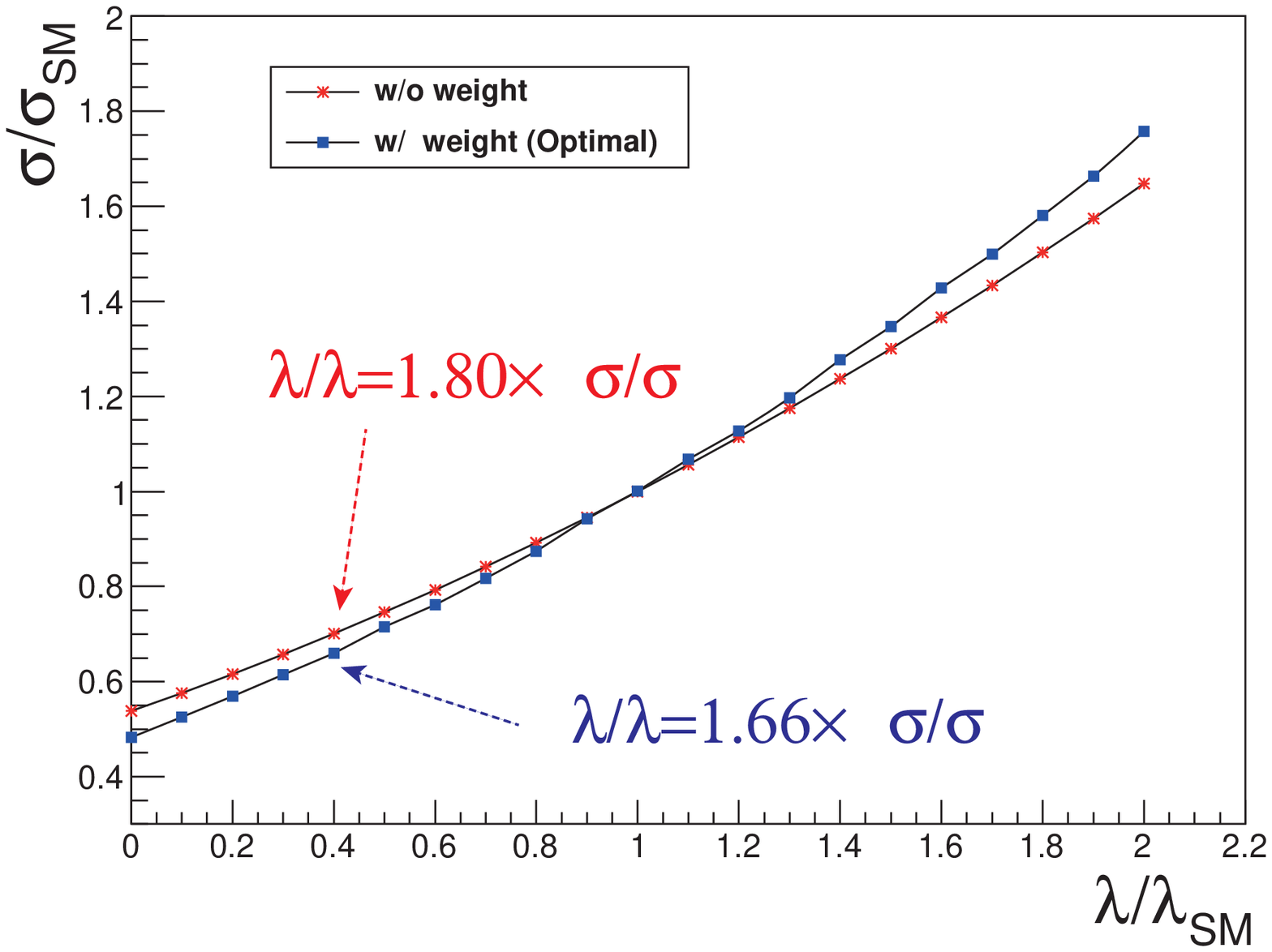}
\hspace{1cm}
\includegraphics[height=50mm]{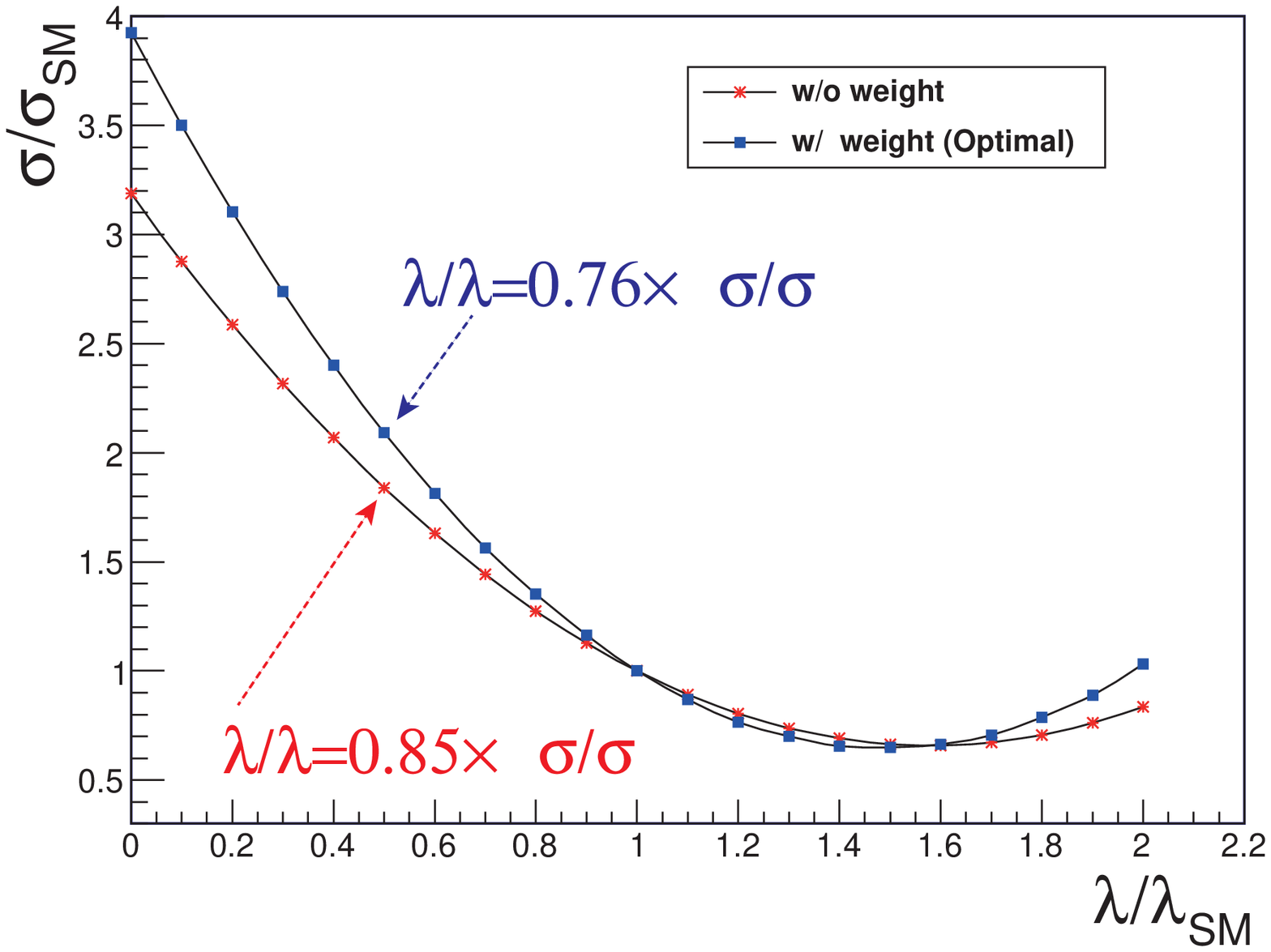}
\caption{Cross sections for the signal $t\bar{t}h$ process with and without the non-relativistic QCD corrections together with those for the background processes: $t\bar{t}Z, t\bar{t}g (g \to b\bar{b})$ and $t\bar{t}$.} 
\label{fig:sensitivity}
\end{figure}

Even more interesting is the measurement of the Higgs self-coupling, since
we need to observe the force that makes the Higgs boson condense in the vacuum
in order to uncover the secret of the EW symmetry breaking.
In other words, we need to measure the shape of the Higgs potential.
There are two ways to measure the self-coupling.
The first method is to use the double higgsstrahlung process: $e^+e^- \to Zhh$
and the second is by the double Higgs production via $W$-fusion: $e^+e^- \to \nu\bar{\nu}hh$.
The first process attains its cross section maximum at around $\sqrt{s}=500\,$GeV,
while the second is negligible there but starts to dominate at energies above $\sqrt{s}\simeq 1.2\,$TeV, as seen in Fig.\ref{fig:sigzhh_vvhh}.
In any case the signal cross sections are very small ($0.2\,$fb or less) and as seen in Fig.\ref{fig:hhdiagrams} irreducible background diagrams containing no self-coupling dilute the contribution from the self-coupling diagram, thereby degrading the sensitivity to the self-coupling, even if we can control the relatively huge SM backgrounds from $e^+e^- \to t\bar{t}, WWZ, ZZ, Z\gamma, ZZZ$, and $ZZh$.
See Fig.\ref{fig:sensitivity} for the sensitivity factors for $e^+e^- \to Zhh$ at $\sqrt{s}=500\,$GeV and $e^+e^- \to \nu\bar{\nu}hh$ at $\sqrt{s}=1\,$TeV, which are 1.66 (1.80) and 0.76 (0.85), respectively, with (without) weighting to enhance the contribution from the signal diagram. 
Notice that if there were no background diagrams, the sensitivity factor would be $0.5$.
The self-coupling measurement is very difficult even in the clean environment of the ILC and requires a new flavor tagging algorithm that precedes jet-clustering, sophisticated neural-net-based data selection, and the event weighting technique \cite{ref:junping}.
The current state of the art for the $Zhh$ data selection is summarized in Table \ref{tab:zhh500}.
\begin{table}[h]
\begin{center}
\caption{The number of remaining events for the three event selection modes: $Zhh \to (\ell\bar{\ell})(b\bar{b})(b\bar{b})$, $ (\nu\bar{\nu})(b\bar{b})(b\bar{b})$, and $ (q\bar{q})(b\bar{b})(b\bar{b})$ and corresponding excess and measurement sensitivities for $m_h=120\,$GeV 
at $\sqrt{s}=500\,$GeV with $2\,$ab$^{-1}$ and $(e^{-}, e^{+})=(-0.8, +0.3)$ beam polarization.}
\begin{tabular}{|c|c|c|c|c|c|}
   \hline
            $\sqrt{s}$ [GeV] & mode & signal & background &
            \multicolumn{2}{c|}{significance} \\
  \cline{5-6}
  & & & & excess & measurement \\
  \hline
  \hline
  500 & $Zhh \to (\ell\bar{\ell})(b\bar{b})(b\bar{b})$ 
         & 3.7 & 4.3 & 1.5$\sigma$ & 1.1$\sigma$ \\
  \cline{3-6}
         & 
         & 4.5 & 6.0 &  1.5$\sigma$ & 1.2$\sigma$ \\
   \hline
  500 & $Zhh \to (\nu\bar{\nu})(b\bar{b})(b\bar{b})$ 
         & 8.5 & 7.9 & 2.5$\sigma$ & 2.1$\sigma$ \\
   \hline
  500 & $Zhh \to (q\bar{q})(b\bar{b})(b\bar{b})$ 
         & 13.6 & 30.7 & 2.2$\sigma$ & 2.0$\sigma$ \\
  \cline{3-6}
         & 
         & 18.8 & 90.6 &  1.9$\sigma$ & 1.8$\sigma$ \\
   \hline
\end{tabular}
\label{tab:zhh500}
\end{center}
\end{table}
Combining all of these three modes, we can achieve $Zhh$ excess significance of $5\sigma$ and measure the production cross section to $\Delta \sigma / \sigma = 27\%$, which translates to $44 (48)\%$ with (without) the event weighting for $m_h=120\,$GeV at $\sqrt{s}=500\,$GeV with $2\,$ab$^{-1}$ and $(e^{-}, e^{+})=(-0.8, +0.3)$ beam polarization \cite{ref:junping}.
The expected precision is significantly worse than that of the cross section because of the background diagrams.
Since the sensitivity factor for the $e^+e^- \to \nu\bar{\nu}hh$ process is much closer to the ideal $0.5$ and since the cross section for this $W$-fusion double Higgs production process increases with the center of mass energy, let us now discuss the measurements at the energy upgraded ILC at $\sqrt{s}=1\,$TeV.

\section{ILC at 1\,TeV}

The $W$-fusion processes become more and more important at higher energies.
Notice also that the machine luminosity usually scale with the center of mass energy.
Combination of these together with the better sensitivity factor allows us to improve the self-coupling measurement significantly at $\sqrt{s}=1\,$TeV, using the $e^+e^- \to \nu\bar{\nu}hh$ process. 
With $2\,$ab$^{-1}$ and $(e^{-}, e^{+})=(-0.8, +0.2)$ beam polarization at $\sqrt{s}=1\,$TeV,
we would be able to determine the cross section for the $e^+e^- \to \nu\bar{\nu}hh$ process to $\Delta \sigma / \sigma = 23\%$, corresponding to the self-coupling precision of $\Delta \lambda / \lambda = 18 (20)\%$ with (without) the event weighting to enhance the contribution from the signal diagram for $m_h=120\,$GeV \cite{ref:junping}.   

At $\sqrt{s}=1\,$TeV, the $e^+e^- \to t\bar{t}h$ process is also near its cross section maximum, making concurrent measurements of the self-coupling and top Yukawa coupling possible.
We will be able to observe the $e^+e^- \to t\bar{t}h$ events with $7.9\sigma$ significance in 8-jet mode and $8.4\sigma$ significance in lepton-plus-6-jet mode, corresponding to 
the relative error on the top Yukawa coupling of $\Delta g_Y(t) / g_Y(t) = 4.0\%$ with $1\,$ab$^{-1}$ and $(e^{-}, e^{+})=(-0.8, +0.2)$ beam polarization at $\sqrt{s}=1\,$TeV for $m_h=125\,$GeV \cite{ref:tth}.

\begin{figure}[h]
\centering
\includegraphics[height=80mm]{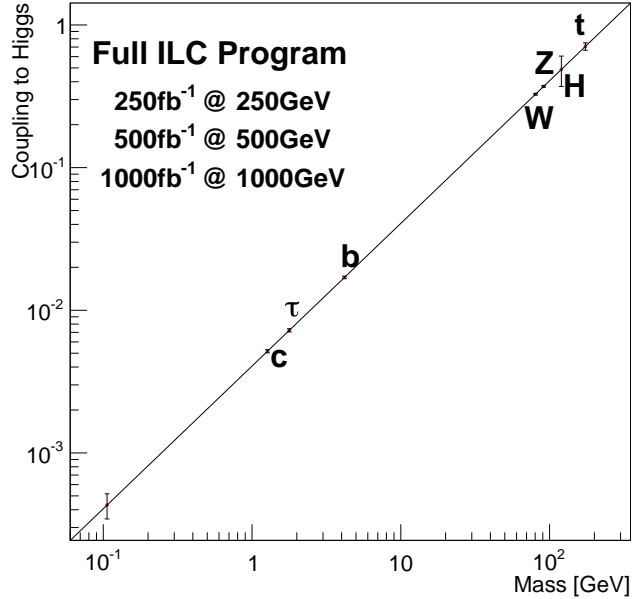}
\caption{Expected mass-coupling relation for the SM case after the full ILC program.} 
\label{fig:mass-coupling-full-ILC}
\end{figure}

Obvious but most important advantage of the higher energy running in terms of Higgs physics is, however, its higher mass reach to the extra Higgs bosons expected in an extended Higgs sector and higher sensitivity to $W_L W_L$ scattering to decide whether the Higgs sector is strongly interacting or not.
In any case thanks to the higher cross section for the $W$-fusion $e^+e^- \to \nu\bar{\nu}h$ process at $\sqrt{s}=1\,$TeV, 
we can expect significantly better precisions for the $\sigma \times BR$ measurements, which allows us to access very rare decays such as $h \to \mu^+\mu^-$ as well as to further improve the precision for the mass-coupling plot (see Fig.\ref{fig:mass-coupling-full-ILC}).

\section{Synergy: LHC + ILC}

\begin{figure}[h]
\centering
\includegraphics[height=65mm]{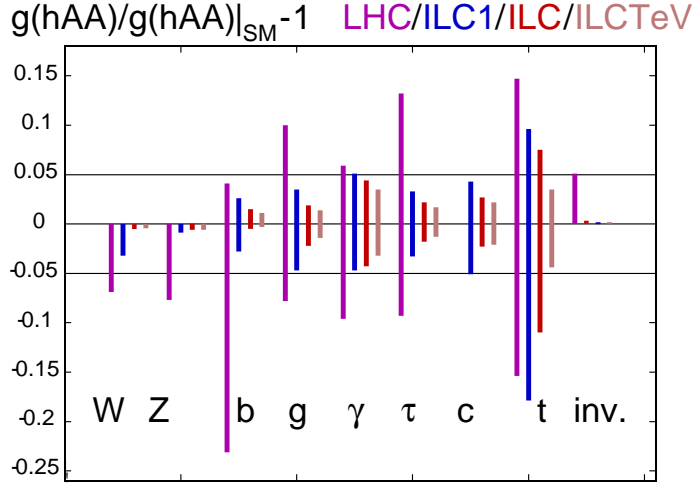}
\caption{Comparison of the capabilities of the LHC and the ILC, when the ILC data in various stages: ILC1 with $250\,$fb$^{-1}$ at $\sqrt{s}=250$, ILC: $500\,$fb$^{-1}$ at $500\,$GeV, and ILCTeV: $1\,$ab$^{-1}$ at $1\,$TeV are cumulatively added to the LHC data with $300\,$fb$^{-1}$ at $14\,$TeV \cite{ref:peskin}.} 
\label{fig:peskin}
\end{figure}

So far we have been discussing the precision Higgs physics expected at the ILC.
It should be emphasized, however, that the LHC is expected to impose significant constraints on possible deviations of the Higgs-related couplings from their SM values by the time the ILC will start its operation, even though fully model-independent analysis is impossible with the LHC alone.
Nevertheless, reference \cite{ref:peskin} demonstrated that with a reasonable weak assumption such as the $hWW$ and $hZZ$ couplings will not exceed the SM values the LHC can make reasonable measurements of most Higgs-related coupling constants except for the $hcc$ coupling.
Figure \ref{fig:peskin} shows how the coupling measurements would be improved by adding, cumulatively, information from the ILC with $250\,$fb$^{-1}$ at $\sqrt{s}=250$, $500\,$fb$^{-1}$ at $500\,$GeV, and $1\,$ab$^{-1}$ at $1\,$TeV to the LHC data with $300\,$fb$^{-1}$ at $14\,$TeV.
The figure tells us that the addition of the $250\,$GeV data, the $hZZ$ coupling in particular, from the ILC allows the absolute normalization and significantly improves all the couplings.
It is interesting to observe the synergy for the measurement of the $h\gamma\gamma$ coupling, whose precision significantly exceeds that of the ILC alone. 
This is because the LHC can precisely determine the ratio of the $h\gamma\gamma$ coupling to the $hZZ$ coupling, while the ILC provides a precision measurement of the $hZZ$ coupling from the recoil mass measurement.
The addition of the $500\,$GeV data from the ILC further improves the precisions, this time largely due to the better determination of the Higgs total width.
Finally as we have seen above, the addition of the $1\,$TeV data from the ILC improves the top Yukawa coupling drastically with even further improvements of all the other couplings except for the $hWW$ and $hZZ$ couplings which are largely limited by the cross section error from the recoil mass measurement at $\sqrt{s}=250\,$GeV. 
This way we will be able to determine these couplings to $O(1\%)$ or better.
The {\it SFitter} group performed a similar but more model-independent analysis and obtained qualitatively the same conclusions \cite{ref:SFitter}.
This level of precision matches what we need to fingerprint different BSM scenarios, when nothing but the 125GeV boson would be found at the LHC (see Table \ref{tab:gupta}).
\begin{table}[h]
\begin{center}
\caption{Maximum deviations when nothing but the 125GeV boson would be found at the LHC \cite{ref:gupta}.}
\begin{tabular}{l c c c}
   \hline
   \hline
   & $\Delta hVV$ & $\Delta h\bar{t}t$ & $\Delta h\bar{b}b$ \\
  \hline
  Mixed-in Singlet & 6\% & 6\% & 6\% \\
  Composite Higgs & 8\% & tens of \% & tens of \% \\
  Minimal Supersymmetry & <1\% & 3\% & 10\%$^a$, 100\%$^b$ \\
  LHC 14\,TeV, 3\,ab$^{-1}$ & 8\% & 10\% & 15\% \\
   \hline
   \hline
\end{tabular}
\label{tab:gupta}
\end{center}
\end{table}
These numbers can be understood from the following formulas for the different models in the decoupling limit \cite{ref:dbd}:
\begin{eqnarray}
\mbox{Mixing with singlet:~~~~~~} & & \cr
	\frac{g_{hVV}}{g_{h_{\rm SM} VV}} = \frac{g_{hff}}{g_{h_{\rm SM} ff}} 
&=& \cos\theta \simeq 1 - \frac{\delta^2}{2} \cr\rule{0in}{5ex}
\mbox{Composite Higgs:~~~~~~~~~} & & \cr
	\frac{g_{hVV}}{g_{h_{\rm SM}VV}} &\simeq& 1 - 3\% 
\left( \frac{1~{\rm TeV}}{f} \right)^2 \cr
	\frac{g_{hff}}{g_{h_{\rm SM}ff}} &\simeq& \left\{ 
	\begin{array}{ll}
	1 - 3\% \left( \frac{1~{\rm TeV}}{f} \right)^2 \qquad& {\rm (MCHM4)} \cr
	1 - 9\% \left( \frac{1~{\rm TeV}}{f} \right)^2 \qquad& {\rm (MCHM5).}
	\end{array} \right. \cr\rule{0in}{3ex}
\mbox{Supersymmetry:~~~~~~~~~~~} & & \cr
	\frac{g_{hbb}}{g_{h_{\rm SM} bb}} &=& 
\frac{g_{h \tau\tau}}{g_{h_{\rm SM} \tau\tau}} \simeq 1 + 1.7\% \left( \frac{1 \ 
{\rm TeV}}{m_A} \right)^2.
\nonumber
\end{eqnarray}
The different models predict different deviation patterns.
The ILC together with the LHC will be able to fingerprint these models or set the lower limit on the energy scale for BSM physics.

\section{Conclusions}
The primary goal for the next decades is to uncover the secret of the electroweak symmetry breaking.
This will open up a window to BSM and set the energy scale for the energy frontier machine that will follow the LHC and the ILC 500.
Probably the LHC will hit systematic limits at $O$(5-10\%) for most of $\sigma \times BR$ measurements, being insufficient to see the BSM effects if we are in the decoupling regime.
To achieve the primary goal we hence need a 500\,GeV linear collider for self-contained precision Higgs studies to complete the mass-coupling plot, where
we start from $e^+e^- \to Zh$ at $\sqrt{s}=250\,$GeV,
then $t\bar{t}$ at around $350\,$GeV,
and then $Zhh$ and $t\bar{t}h$ at $500\,$GeV.
The ILC to cover up to $\sqrt{s}=500\,$GeV is an ideal machine to carry out this mission (regardless of BSM scenarios) and we can do this with staging starting from $\sqrt{s}\simeq250\,$GeV.
We may need more data at this energy depending on the size of the deviation, since the recoil mass measurement eventually limits the coupling precisions.
Luminosity upgrade possibility should be always kept in our scope.
If we are lucky, some extra Higgs boson or some other new particle might be within reach already at the ILC 500. 
Let's hope that the upgraded LHC will make another great discovery in the next run from 2015.
If not, we will most probably need the energy scale information from the precision Higgs studies.
Guided by the energy scale information, we will go hunt direct BSM signals, if necessary, with a new machine. Eventually we will need to measure $W_L W_L$ scattering to decide if the Higgs sector is strongly interacting or not.

In this talk I have been focusing on the case where $X(125)$ alone would be the probe for BSM physics, but there is a good chance for the higher energy run of the LHC to bring us more.
It is also very important to stress that the ILC, too, is an energy frontier machine.
It will access the energy region never explored with any lepton collider before.
There can be a zoo of new uncolored particles or new phenomena that are difficult to find at the LHC but can be discovered and studied in detail at the ILC.
For instance, natural SUSY where the $\mu$ parameter not far above $100\,$GeV, we expect relatively light chargino and neutralinos which are higgsino-dominant and hence nearly mass-degenerate (typically $\Delta m$ of a few GeV or less), a very difficult case for the LHC.
At the ILC $\Delta m$ as small as $50\,$MeV can be handled with the ISR tagging.
If $\Delta m=400\,$MeV or so, we can determine the masses to $2\,$GeV and $\Delta m$ to $7\,$MeV.
If this is the case, the ILC will be not only the Higgs factory but also a Higgsino factory \cite{ref:dbd}.
Another example is search for possible anomalies in precision studies of properties of $W/Z$ and top, or two-fermion processes \cite{ref:dbd}.
Whatever new physics awaits us, clean environment, polarized beams, and excellent jet energy resolution to reconstruct $W, Z, t,$ and $h$ in their hadronic decays will enable us to uncover the nature of the new physics through model-independent precision measurements.

\begin{acknowledgments}
The materials presented in this talk were prepared for the ILC TDR physics chapter in collaboration with the members of the ILC physics working group \cite{ref:ilcphys} and the members of the ILC physics panel. 
The author would like to thank them for useful discussions, especially M.\,Peskin, Y.\,Okada, S.\,Kanemura., J.\,Tian, H.\,Ono, and T.\,Tanabe. This work is supported in part by the Creative Scientific Research Grant No.\,18GS0202 of the Japan Society for Promotions of Science (JSPS), the JSPS Grant-in-Aid for Science Research No.\,22244031, and the JSPS Specially Promoted Research No.\,23000002.
\end{acknowledgments}

\bigskip 

\end{document}